\documentclass{PoS}
\usepackage{amssymb,amsmath,amsfonts}
\usepackage{graphicx}
\graphicspath{{./Figs/}}
\usepackage{slashed}

\newcommand{\Tr}{\operatorname{Tr}}
\newcommand{\Eq}[1]{Eq.\,\eqref{#1}}

\newcommand{\Tab}[1]{Table\,\ref{#1}}
\newcommand{\RGI}{\mathsf{RGI}}

\title{Towards lattice-assisted hadron physics calculations based on QCD n-point functions}
\ShortTitle{Towards lattice-assisted hadron physics calculations based on QCD n-point functions}

\author{\speaker{Andr\'e Sternbeck}\\\\
        Theoretisch-Physikalisches Institut,
        Friedrich-Schiller-Universit\"at Jena, 07743 Jena, Germany\\
        E-mail: \email{andre.sternbeck@uni-jena.de}}

\author{Milena Leutnant\\
        Theoretisch-Physikalisches Institut,
        Friedrich-Schiller-Universit\"at Jena, 07743 Jena, Germany}

\author{Gernot Eichmann\\
        CFTP, Instituto Superior Tecnico, Universidade de Lisboa, 1049-001 Lisboa, Portugal}

\abstract{We present preliminary lattice results for the nonperturbative tensor structure of the vector and axial-vector quark-antiquark vertices in QCD. Our lattice calculations are for $N_f=2$ mass-degenerate Wilson fermion flavors whose quark mass values include an almost physical one. We compare our lattice results with the corresponding continuum solutions of the inhomogeneous Bethe-Salpeter equations in a rainbow-ladder truncation. We find similarities in the momentum dependencies of the form factors but also clear deviations at low momentum.}

\FullConference{The 36th Annual International Symposium on Lattice Field Theory --- LATTICE2018\\
		22--28 July, 2018, 
		Michigan State University, East Lansing, Michigan, USA.}

\begin{document}

\section{Motivation}

Lattice QCD calculations are currently the preferred first-principle tool for hadron physics. Euclidean two-point correlation functions give us access to the hadron spectrum and three-point correlation functions to matrix elements and the associated hadronic form factors. In principle, the lattice regularization offers full control over all systematics errors, although in practice this is often hard to achieve as the numerical effort may grow tremendously.

Lattice calculations are not the only way to address strong-interaction physics. Numerical calculations based on functional
methods such as Dyson-Schwinger and Bethe-Salpeter equations (DSEs and BSEs) form another approach,
which can be useful in constraining quantities also beyond the space-like domain, see e.g.~\cite{Eichmann:2016yit}.
However, since this approach builds upon an infinite tower of relations between quark and gluon $n$-point functions which has to be truncated for numerical treatments,
it comes with a systematic error that is difficult to quantify without additional input such as from lattice QCD.

\begin{floatingfigure}[r]
 \includegraphics[width=4.3cm]{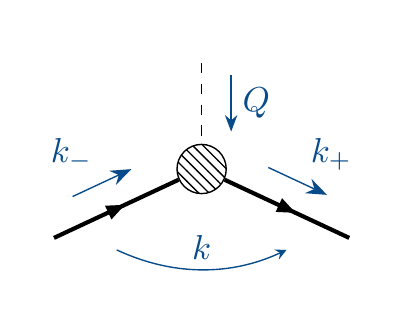}
 \caption{Quark-bilinear current coupled to external particle (e.g., photon) with momentum $Q$.
  \label{fig:qbqvertex}}
\end{floatingfigure}
To reliably estimate truncation errors, lattice QCD input should not be restricted to hadronic quantities but
also include quark and gluon $n$-point functions. Lattice studies in the past focused on quark and gluon propagators or the triple-gluon and quark-gluon vertex functions of QCD. To access hadronic form factors within functional methods, the offshell tensor structure of quark-bilinear operators is required as well. For instance, the non-perturbative dressing of the quark-photon vertex associated with the vector current is an essential ingredient for electromagnetic or transition form factor calculations. The corresponding quark-bilinear (current) 3-point functions are typically calculated on the lattice when targeting RI'(S)MOM renormalization constants. There, however, not the complete tensor structure of the underlying vertex is determined but rather the divergent terms at a fixed scale.

We calculate, for the first time, the QCD dressing for all form factors of the vector and axial-vector quark-antiquark vertices for one off-shell kinematic configuration on the lattice. In the continuum, these vertices have the general decomposition
\begin{align}
 \label{eq:GammaV}
 \Gamma_V^\mu(k,Q) &= g_1^V \gamma^\mu + g_2^V k^\mu\slashed{k} + g_3^V ik^\mu
            + g_4^V\frac{i\omega}{2}[\gamma^\mu,\slashed{k}] \nonumber \\
            &\quad + f_1^V T^\mu_1 + f_2^V\omega T^\mu_2 + f_3^V T^\mu_3 + f_4^V T^\mu_4 + f_5^V T_5^\mu + f_6^V T^\mu_6 + f_7^V\omega T^\mu_7 + f_8^V T_8^\mu \,,
    \\ 
 \label{eq:Gamma5V}
 \gamma_5\Gamma_A^\mu(k,Q) &= g_1^A\gamma^\mu + g_2^A k^\mu\slashed{k} + g_3^AiQ^\mu
            + g_4^A\frac{i}{2}[\gamma^\mu,\slashed{k}] \nonumber \\
            &\quad + f_1^A T^\mu_1 + f_2^A T^\mu_2
            + f_3^A\omega T^\mu_3 + f_4^A T^\mu_4  + f_5^A\frac{\omega}{Q^2} T_5^\mu + f_6^A T^\mu_6 + f_7^A\omega T^\mu_7 + f_8^A\omega T_8^\mu \,,
\end{align}
where the tensors $T_{1\dots 8}^\mu$ are given by
             \begin{equation*}  \renewcommand{\arraystretch}{1.3}
             \begin{array}{r@{\!\;\,}l   }

                 T_1^\mu &= t^{\mu\nu}_{QQ}\,\gamma^\nu,  \\
                 T_2^\mu &= t^{\mu\nu}_{QQ}\, \tfrac{i}{2} [\gamma^\nu,\slashed{k}]\,,

             \end{array} \qquad
             \begin{array}{r@{\!\;\,}l   }

                 T_3^\mu &= \tfrac{i}{2}\,[\gamma^\mu,\slashed{Q}] \,, \\
                 T_4^\mu &= \tfrac{1}{6}\,[\gamma^\mu, \slashed{k}, \slashed{Q}]\,,

             \end{array} \qquad
             \begin{array}{r@{\!\;\,}l   }

                  T_5^\mu &= t^{\mu\nu}_{QQ}\,ik^\nu,  \\
                  T_6^\mu &= t^{\mu\nu}_{QQ}\,k^\nu \slashed{k}\,,

             \end{array} \qquad
             \begin{array}{r@{\!\;\,}l   }

                  T_7^\mu &= t^{\mu\nu}_{Qk}\,\gamma^\nu,  \\
                  T_8^\mu &= t^{\mu\nu}_{Qk}\,\tfrac{i}{2}\,[\gamma^\nu,\slashed{k}] 

             \end{array}
             \end{equation*}
and we used $t^{\mu\nu}_{AB} = A\cdot B\,\delta^{\mu\nu} - B^\mu A^\nu$
and $\tfrac{1}{6}\,[\gamma^\mu,\gamma^\nu,\gamma^\rho] = -\gamma_5\,\varepsilon^{\mu\nu\rho\sigma}\gamma^\sigma$.
Here, $Q$ is the incoming momentum (e.g., of the virtual photon for the case of $\Gamma_V$) and $k$ is the relative quark momentum. The outgoing and incoming quark and anti-quark have momenta $k_\pm = k \pm Q/2$, respectively (see Fig.\ref{fig:qbqvertex}) and we denoted $\omega=k\cdot Q$.
The above decompositions with gauge parts constructed from the form factors $g_i(k^2,\omega,Q^2)$
and transverse parts from $f_i(k^2,\omega,Q^2)$ follow from implementing the vector and axialvector Ward-Takahashi identities (WTIs) without introducing kinematic singularities.
As a result, the $g_i$ and $f_i$ are even in the variable $\omega$ and non-singular in the kinematic limits $k^\mu\to 0$ and $Q^\mu \to 0$. In the vector case, the first line in \Eq{eq:GammaV}
is the Ball-Chiu vertex~\cite{Ball:1980ay}.

\section{Lattice setup}

\begin{table*}
\centering
\begin{tabular}{l@{\;\;\;}c@{\;\;\;}c@{\;\;\;}c@{\;\;\;}c@{\;\;\;}c@{\;\;}c@{\;\;}c@{\;\;}c@{}r@{\quad}}
 no. &$\beta$ & $\kappa$  & $V$ & $a$\,[fm] & $Z^\RGI_\psi$ & $Z^\RGI_V$ & $Z^\RGI_A$ & $m_\pi$\,[MeV] &  \multicolumn{1}{c}{\#cnfg.} \\ \hline\hline
C-1 & 5.20 & 0.13550 & $32^3\!\times64$ & 0.081 & 0.749 & 0.722 & 0.753 & $681$  & 8 \\
C-3 &  & 0.13584 & $32^3\!\times64$ &  &&&& $409$  & 8 \\*[0.7ex]
E-3 & 5.29 & 0.13620 & $32^3\!\times64$ & 0.071 & 0.759 & 0.737 & 0.765 & $422$ & 10 \\
E-4  &  & 0.13632  & $64^3\!\times64$ & &&&& $290$  & 30\\
E-5 &  & 0.13640   & $64^3\!\times64$ &  &&&& $150$  & 238 \\*[0.7ex]
F-3 & 5.40 & 0.13647  & $32^3\!\times64$ & 0.060 & 0.771 & 0.751 & 0.778 & $426$  & 10 \\
\hline \hline
\end{tabular}
\caption{Lattice parameters for our gauge ensembles as given by the RQCD collaboration \cite{Bali:2016lvx}. The RGI renormalization constants $Z^\RGI$ are updates of the values in \cite{Gockeler:2010yr}; they correspond to $r_0=0.5\,\text{fm}$ and $r_0\Lambda^{\overline{\mathsf{MS}}}=0.789$. The values for $c_{SW}$ are $2.0171$, 1.9192 and 1.8228 for $\beta=5.20$, 5.29 and 5.40, respectively \cite{Bali:2016lvx}. For the $64^4$ lattices the statistics varies with momentum and \#cnfg.\ refers here to the number of analyzed gauge configurations for the lowest five momenta (see text). For higher momenta smaller amounts are sufficient.}
\label{tab:stat}
\end{table*}

For our lattice calculation we use the $N_f=2$ gauge ensembles of the RQCD collaborations (Wilson action, clover-improved fermions). We gauge fix a subset (see \Tab{tab:stat}) to Landau gauge  and calculate the quark propagator $S$ and the connected part of the quark-bilinear 3-point functions:
\begin{align}
 S(k_\pm) &= \frac{1}{V}\sum_{x,y} e^{ik_\pm(x-y)} \left\langle \left[D_W^{-1}(U;x,y)\right] \right\rangle_U \,,
  \label{eq:Sonlattice}
\\ 
 G_\Lambda(k,Q) &= \frac{1}{V^{3/2}}\sum_{x,y,z} e^{ik_+(x-z) + ik_-(z-y)} \left\langle D_W^{-1}(U;x,z)\,\Lambda\,
D_W^{-1}(U;z,y)\right\rangle_U \,.
  \label{eq:GLambda}
 \end{align}
 $\Lambda$ is any of the Dirac matrices $\Lambda=\gamma_\mu,\gamma_5\gamma_\mu$ and $D_W$ denotes the Wilson-clover fermion matrix. To have an optimal signal-to-noise ratio we use the plane-wave-source method for the inversion of $D_W$. The vertex is obtained form the amputated 3-point function,
 \begin{equation}
 \Gamma_\Lambda(k,Q) = S^{-1}(k_+)\:G_\Lambda(k,Q)\:S^{-1}(k_-)\,,
 \label{eq:Gmu_amputated}
\end{equation}
and its form factors from projecting $\Gamma_\Lambda$ onto its tensor structure. For example, for the vector vertex this yields
\begin{equation}
    \label{eq:ffinv}
     \{ g_i, f_i\} = \sum_i\left[\mathbf{P}^{-1}\right]_{ij} s_j \qquad\text{with the traces}\quad s_j = \Tr\left\{\Gamma_V^\mu(k,Q)\,P_j^\mu(k,Q) \right\}
\end{equation}
and the matrix elements $\left[\mathbf{P}(k,Q)\right]_{ij} = \Tr\big\{P_i^\mu(k,Q)\, P_j^\mu(k,Q)\big\}$. Here $P_i^\mu(k,Q)$ refers to one of the base tensors of $\Gamma_V^\mu$ in \Eq{eq:GammaV}. 

For the quark propagator momenta $k_\pm$ we choose an \emph{asymmetric} setup ($Q^2 = k_+^2\neq k_{-}^2$) and in addition we use twisted boundary conditions for the fermions. This adds a shift to the momenta proportional to the twist angle (see, e.g., \cite{Arthur:2010ht}):
\begin{equation}
  ak^\mu_{\pm} =  \frac{2\pi}{N_\mu} \left(n^\mu_\pm + \frac{\tau^\mu_\pm}{2}\right),
 \label{eq:momenta}
\end{equation}
 which we exploit to enhance the momentum resolution. For our choice of momenta the integer components, $n^\mu_\pm$, and twist angles, $\tau^\mu_\pm$, for the respective $k_-$ and $k_+$ read
\begin{align}
  n_+ &= n\,(2,1,0,0),\qquad n_- = n\,(0,1,1,0)\qquad\text{with}\quad n=1,2,\ldots,N_s/4\;. \\
  \tau_+ &= \tau\,(2,1,0,0),\qquad \tau_- = \tau\,(0,1,1,0)\qquad\text{with}\quad \tau=0,0.4, 0.8, 1.2\ \text{and}\ 1.6.
\end{align}
This corresponds to
\begin{equation}
  Q^2 = k_+^2 = \frac{5}{2}k^2_-\:, \qquad k^2 = \frac{9}{20}Q^2 \qquad\text{and}\qquad \zeta^2 \equiv \frac{\omega^2}{k^2Q^2}=\frac{1}{5}\,.
\end{equation}

Our quark bilinears and propagator are not offshell $O(a)$-improved; only the action is correct to $O(a)$. We thus expect that the form factors will deviate from their behavior in the continuum for higher $k_\pm^2$, and at small momentum where the deviations increase with the bare quark mass [$O(am)$ effects]. Statistical errors are estimated with a Bootstrap analysis. We find they are drastically enhanced at small momentum for the smallest (almost physical) quark mass; occasional outliers appear in the Monte-Carlo history for ensemble E-5. For the $64^4$ lattices the number of configurations is therefore enlarged for the lowest five $k_\pm^2$. The numbers quoted for E-4 and E-5 in \Tab{tab:stat} (\#cnfg.) refer to the lowest five momenta.\footnote{Due to the twisted boundary condition 5-tuples of adjacent momenta belong to the same $n_\pm$ but differ by $\tau_\pm$.} For higher momenta a lower number is sufficient: 14 configurations for ensemble E-4 and 64 for ensemble E-5, with the exception for the second lowest momentum 5-tuple of ensemble E-5 where 160 gauge field configurations are analyzed.

\section{Lattice results for the form factors}

\begin{figure*}[t]\centering
\includegraphics[width=0.43\textwidth]{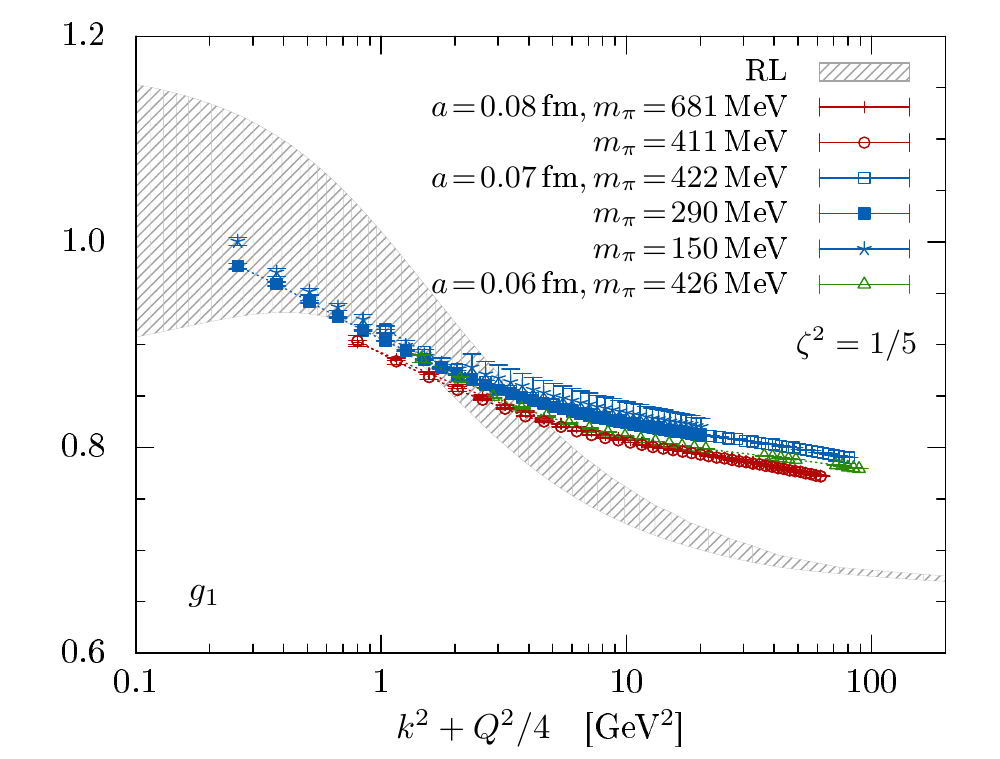}
\includegraphics[width=0.43\textwidth]{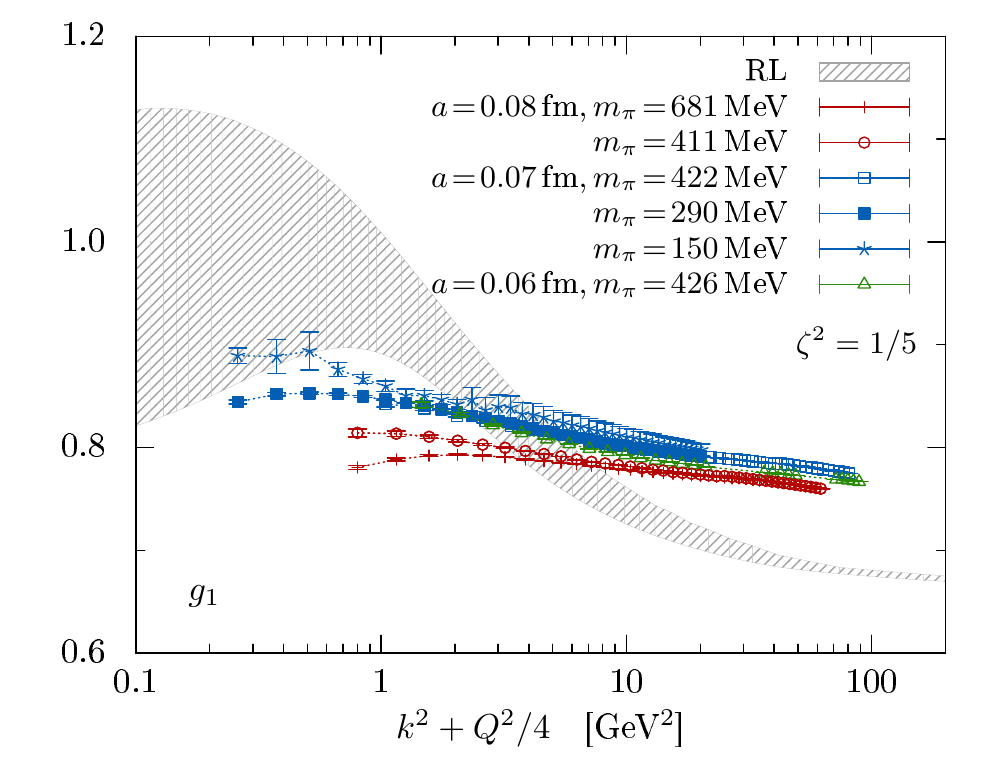}
\includegraphics[width=0.43\textwidth]{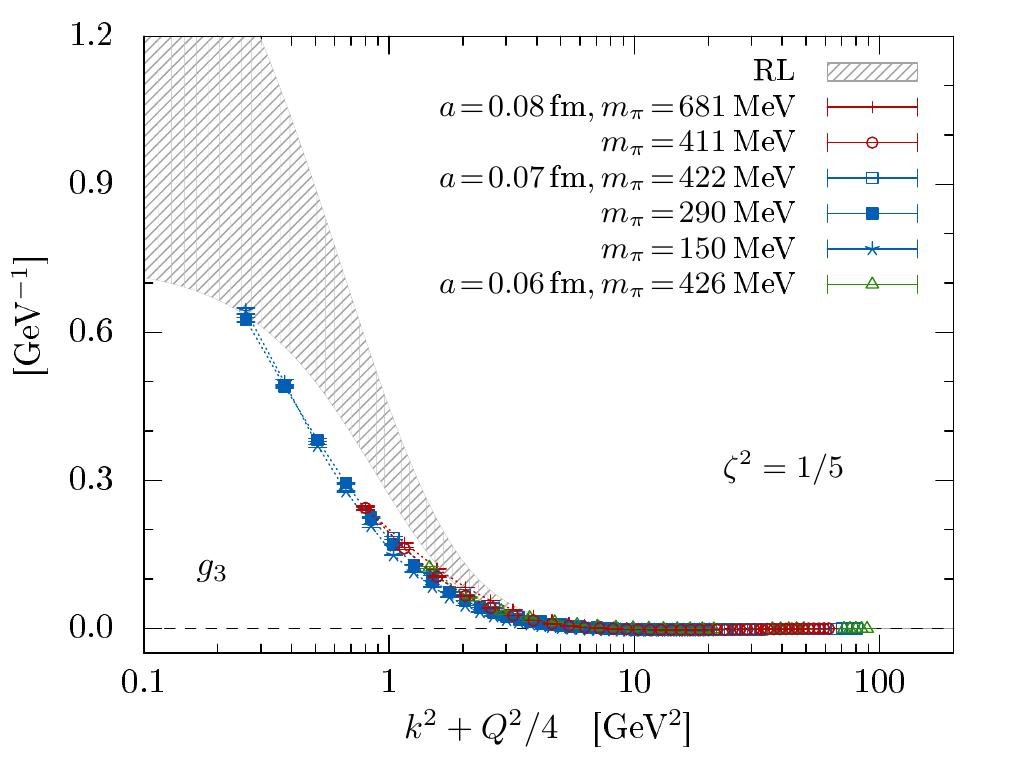}
\includegraphics[width=0.43\textwidth]{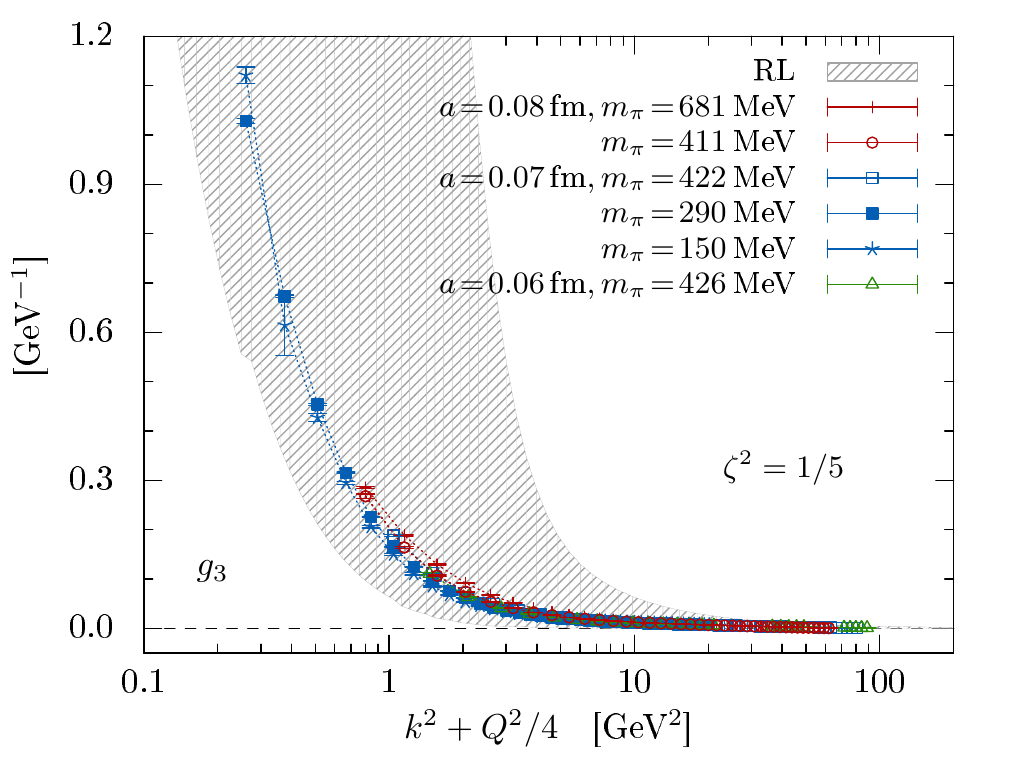}
\caption{Renormalized gauge form factors of the vector (left) and axial-vector vertex (right). \label{fig:qpv_axv_gauge_ff}}
\end{figure*}

\begin{figure*}\centering
\includegraphics[width=0.45\textwidth]{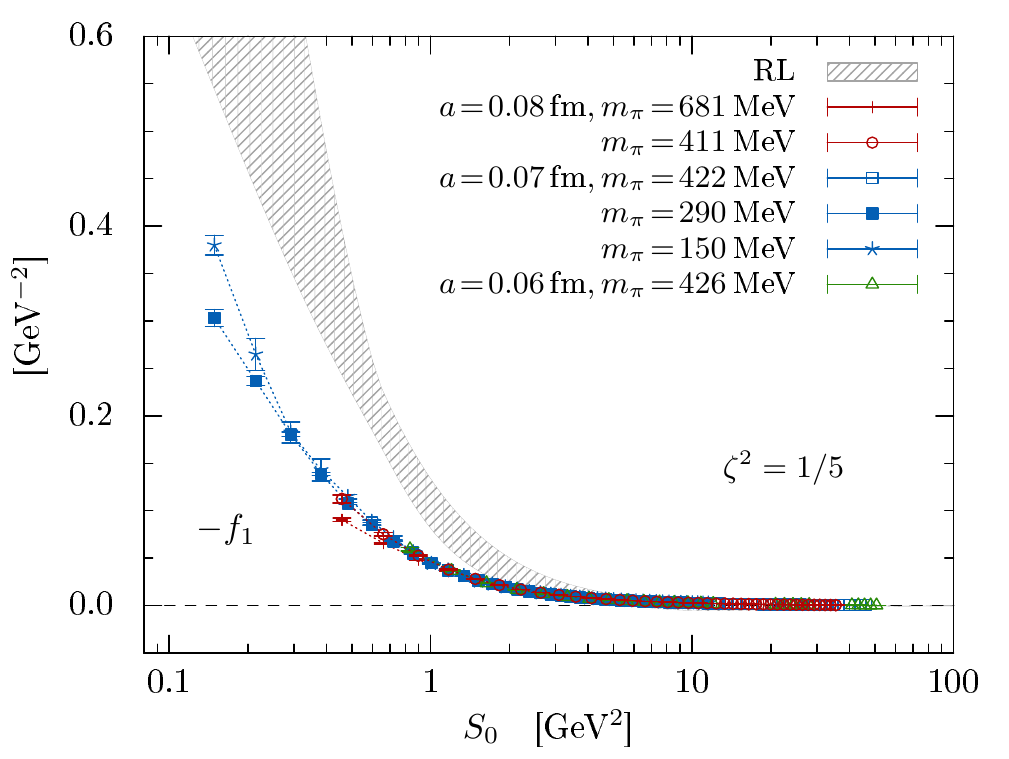}
\includegraphics[width=0.45\textwidth]{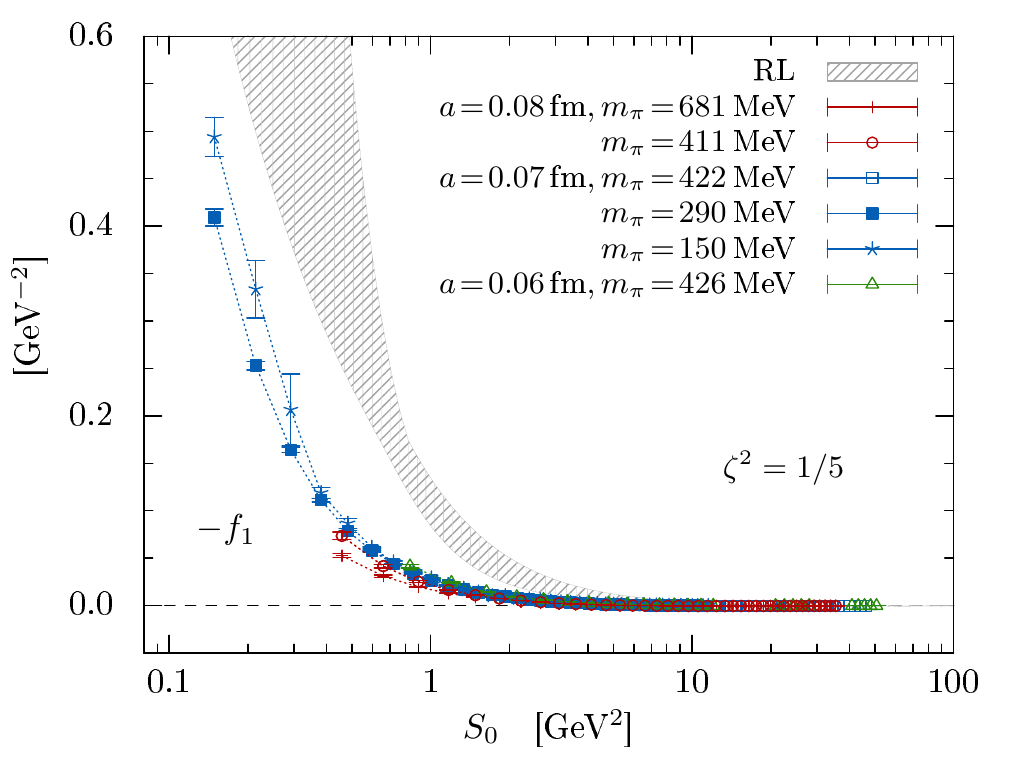}
\includegraphics[width=0.45\textwidth]{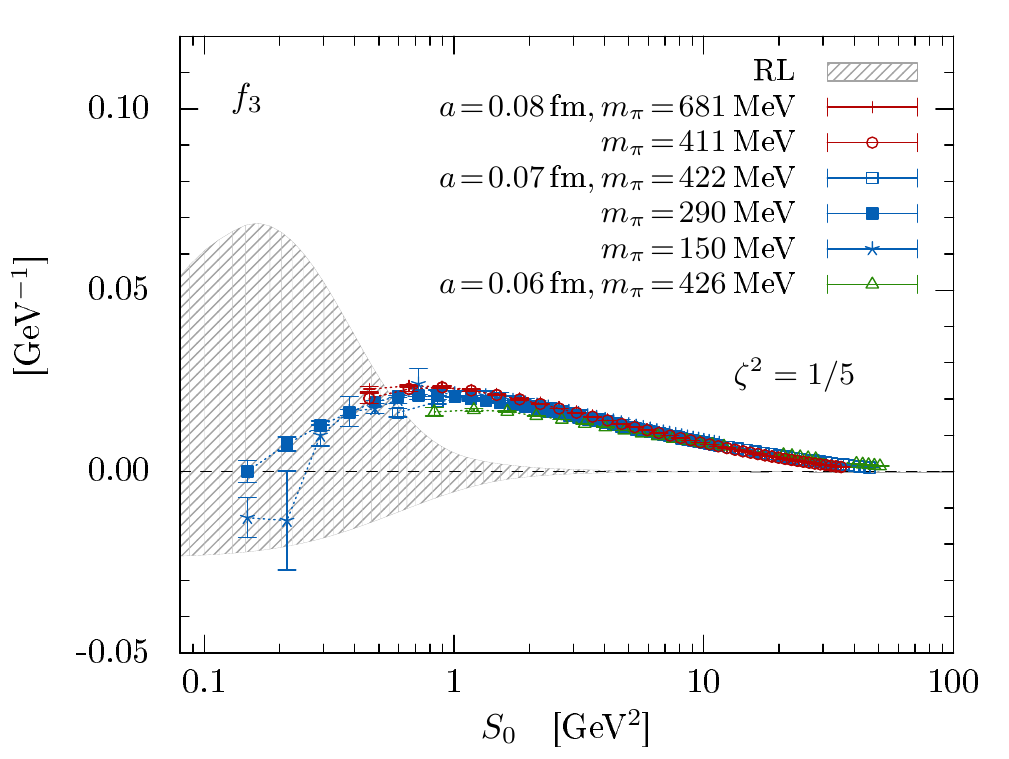}
\includegraphics[width=0.45\textwidth]{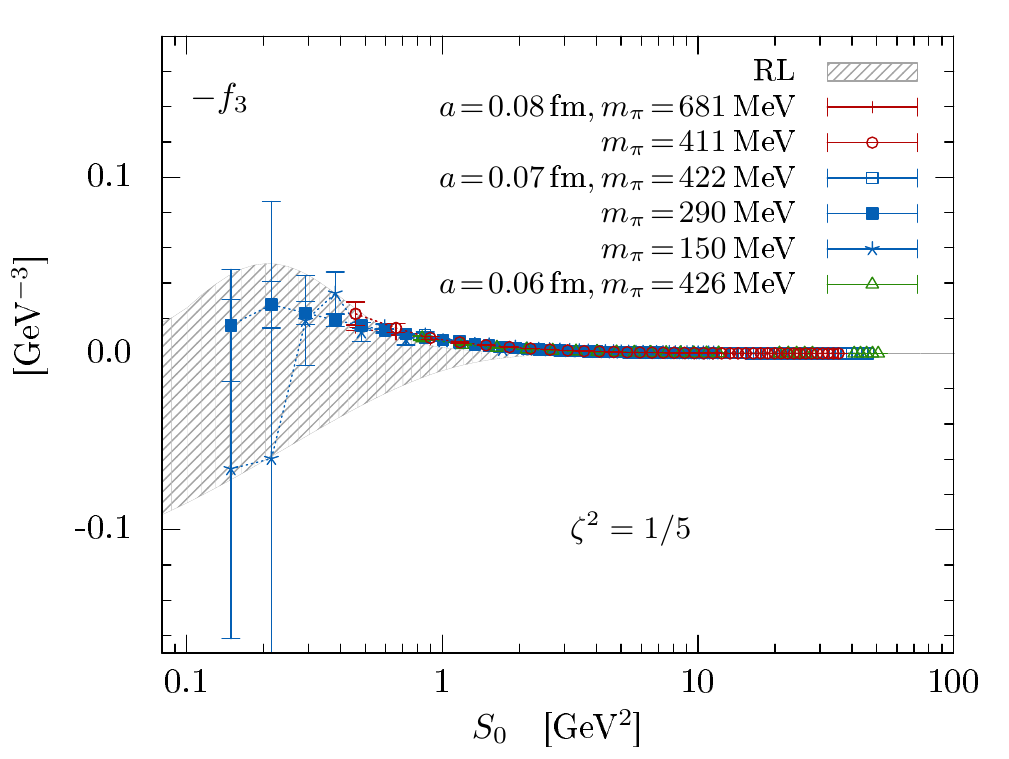}
\includegraphics[width=0.45\textwidth]{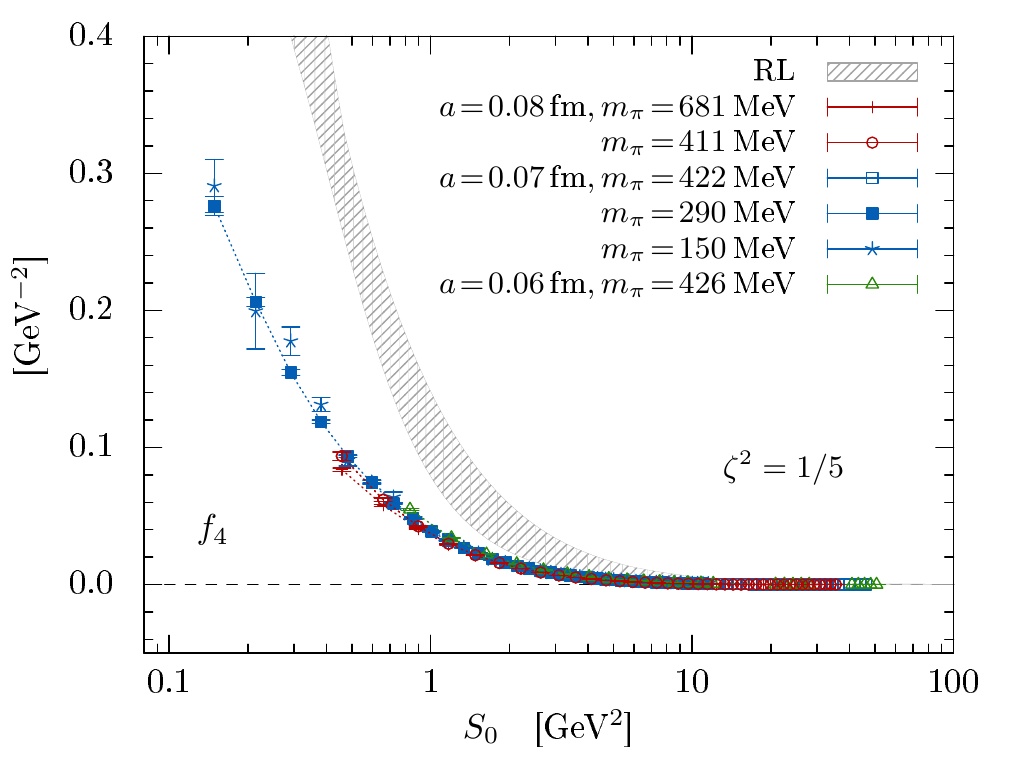}
\includegraphics[width=0.45\textwidth]{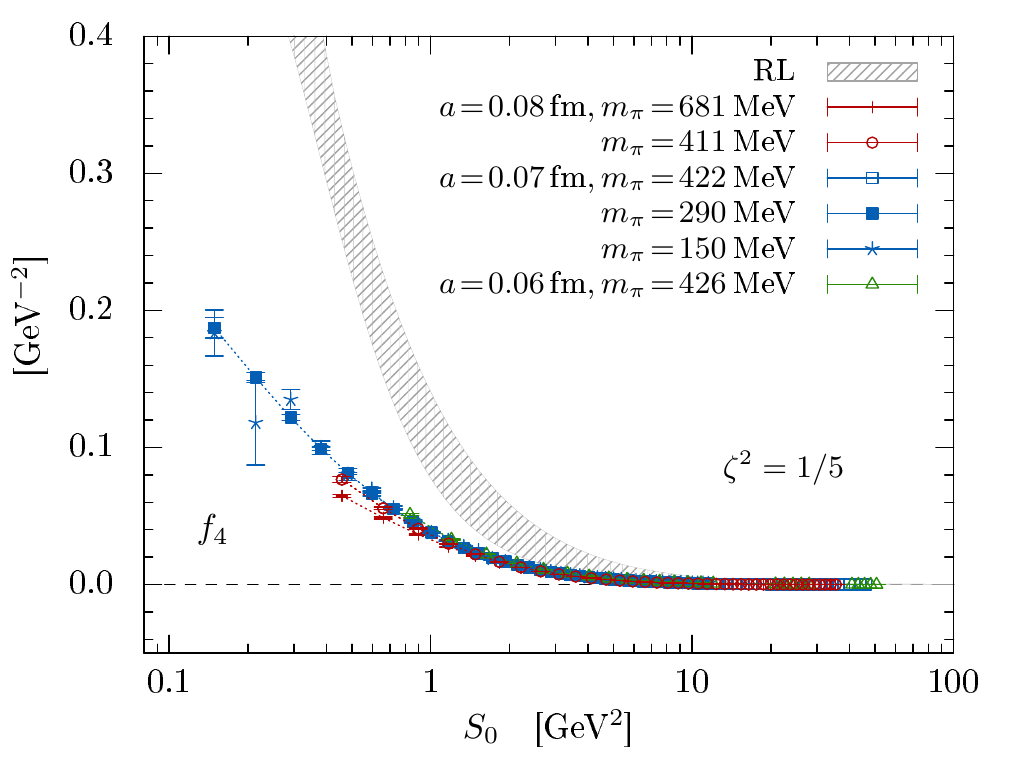}
\includegraphics[width=0.45\textwidth]{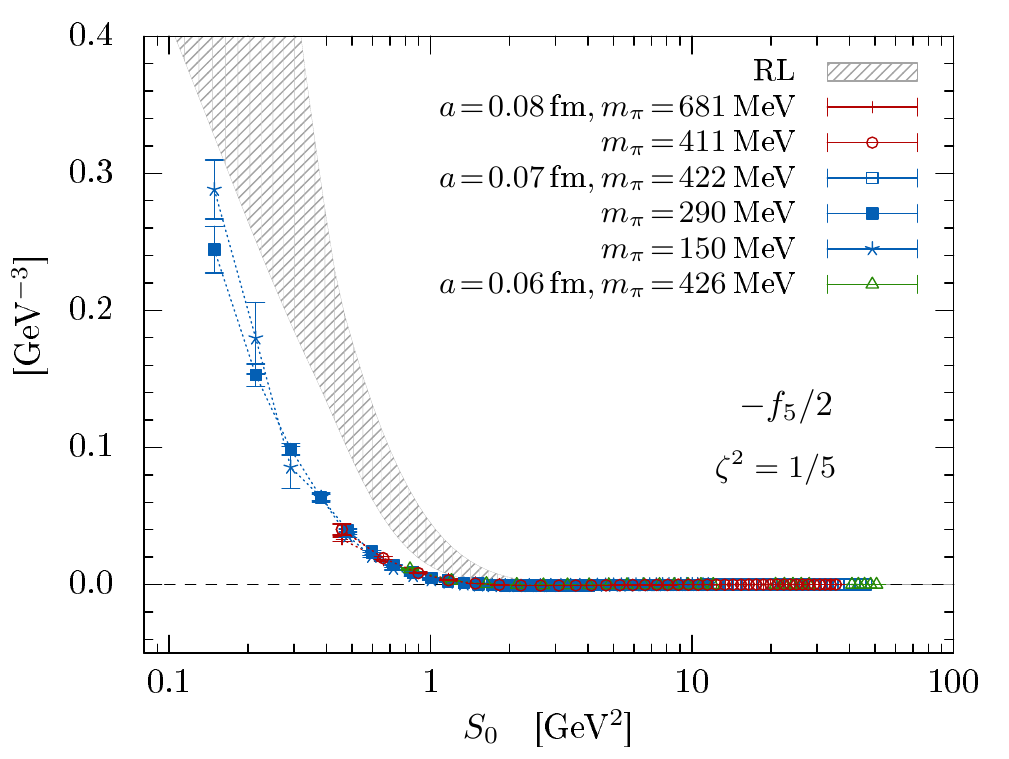}
\includegraphics[width=0.45\textwidth]{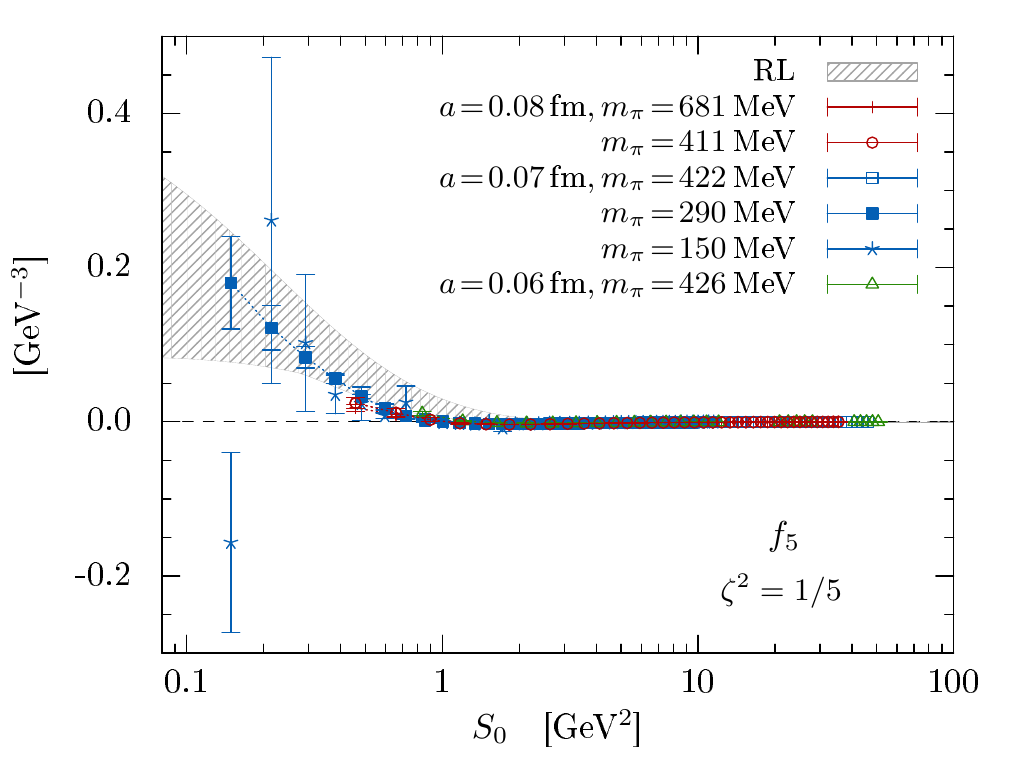}
\caption{Renormalized transverse form factors of the vector (left) and axial vector vertex (right). \label{fig:qpv_axv_transf}}
\end{figure*}

We calculate all vertex form factors on the ensembles listed in \Tab{tab:stat}. 
Selected lattice results are presented in Figs.\,\ref{fig:qpv_axv_gauge_ff} and \ref{fig:qpv_axv_transf} and additional ones can be found in~\cite{Leutnant:2018dry}
and in the forthcoming publications. The results shown here are renormalized with the RGI renormalization constants of the RQCD collaboration given in \Tab{tab:stat}, i.e., $\Gamma_{\Lambda}^R = Z_\Lambda Z^{-1}_\psi\Gamma_\Lambda$ (see, e.g., \cite{Gockeler:2010yr} for details).\footnote{Note that small corrections to these chirally-extrapolated constants would yield a better overlap for the $g_1^{V,A}$ data.}

To compare with solutions of the rainbow-ladder truncated inhomogeneous BSEs, 
we plot the form factors $g_i$ versus $k^2+Q^2/4$ and $f_i$ versus $S_0 \equiv Q^2/4 + k^2/3$. 
These capture the main momentum dependencies, i.e., the form factors $g_i(k^2,\omega,Q^2)$  from the BSE solutions mainly scale with $k^2+Q^2/4$ and the $f_i(k^2,\omega,Q^2)$ with $S_0$.
For the lattice momentum setup these variables reduce to $k^2+Q^2/4 = 7Q^2/10$ and $S_0=2Q^2/5$, respectively. 
The BSE bands in Figs.\,\ref{fig:qpv_axv_gauge_ff} and \ref{fig:qpv_axv_transf} 
correspond to zeroth Chebyshev moments in the variable $\zeta = -1 \dots 1$ and
contain the full spacelike kinematic dependencies for $k^2>0$ and $Q^2>0$. 
In addition, they absorb the rainbow-ladder model dependence by varying the parameter $\eta$ from $1.6 \dots 2.0$, cf.~\cite{Eichmann:2016yit} for details.
The remaining scale parameter and current-quark mass were chosen to reproduce the pion decay constant 
at the physical pion mass $m_\pi=138$ MeV. The continuum solutions were obtained
in a MOM scheme, where due to multiplicative renormalizability the renormalization constants $Z_\psi = Z_V = Z_A$ are identical. To match with the lattice results, we have renormalized all form factors
such that $g_1$ for the vector vertex agrees with the lattice data at $k^2+Q^2/4 = 3$ GeV$^2$ for the central value of $\eta$.

In general the continuum solutions and lattice data show a similar momentum dependence.
There are, however, clear deviations at low momenta, where the
lattice results show a milder momentum dependence and
zooming in further reveals $O(am)$ effects at low momentum.

The form factors encode the information about how an external current couples to a non-perturbative quark.
For example, the vector and axial-vector WTIs entail that $g_1^V = g_1^A$ in the chiral limit,
where both are completely determined by the quark propagator. This trend can be seen in Fig.~\ref{fig:qpv_axv_gauge_ff}
(the deviations at large momenta are likely lattice artifacts~\cite{Leutnant:2018dry}).
The axial WTI entails that $g_3^A$ encodes the pion pole in the timelike region, i.e., $g_3^A \sim 1/(Q^2+m_\pi^2)$,
which explains the strong rise in the infrared.
The form factors $f_i$ in Fig.~\ref{fig:qpv_axv_transf} contain further timelike poles because a photon, $W$ or $Z$ boson can fluctuate into particles
with matching quantum numbers. In the timelike region $f_1^V$ must have vector-meson poles with quantum numbers $J^{PC} = 1^{--}$, 
whereas $f_1^A$ encodes again the pion pole but also axial-vector meson poles.
If $\Gamma_V^\mu(k,Q)$ is taken onshell and contracted with Dirac spinors, one obtains the 
onshell matrix element with Dirac and Pauli form factors $F_1(Q^2)$ and $F_2(Q^2)$, where both 
$f_3^V$ and $f_4^V$ contribute to the anomalous magnetic moment $F_2(0)$. In Fig.~\ref{fig:qpv_axv_transf} one can see 
that the overall smallness of the quark anomalous magnetic moment $f_3^V$
does not appear to be a deficiency of the rainbow-ladder truncation but rather a genuine feature in QCD.

\newpage

\section{Summary}

We have performed the first lattice calculation of the nonperturbative tensor structure of the vector and axial-vector vertices. Our calculation is for $N_f=2$ mass-degenerate Wilson fermions and expands over three lattice spacings and several bare quark masses; disconnected contributions are not yet included though. When comparing lattice results to
the available continuum solutions (rainbow-ladder truncation), we find that the form factors have a similar momentum dependence but there are clear deviations towards low momenta. Lattice spacing artifacts are surprisingly small but seen when zooming in. For instance, a check of the vector WTI with data for an offshell $O(a)$-improved quark propagator reveals that for $Q^2>4\,\text{GeV}^2$ lattice spacing artifacts become important (see \cite{Leutnant:2018dry}). 
Future lattice studies should use offshell $O(a)$-improved Wilson fermions (see \cite{Heatlie:1990kg,Skullerud:2000un,Oliveira:2018lln}) and use point-split currents \cite{Karsten:1980wd,Horsley:2015nae}. This will help to improve our lattice results in the transition regime to perturbation theory and reduce $O(am)$ effects at small momentum.

%
{\medskip\footnotesize
We thank the RQCD collaboration for their gauge configurations. The gauge fixing and calculations of fermion propagators were performed on the HLRN supercomputing facilities (Berlin/Hannover), the Ara cluster of the FSU Jena and the Leibniz Supercomputing Center of the Bavarian Academy of Sciences and Humanities (LRZ) on the supercomputer SuperMUC. 
This work was supported by the BMBF under grant No.\ 05P15SJFAA (FAIR-APPA-SPARC), the DFG Research Training Group GRK1523, and the FCT Investigator Grant IF/00898/2015.
}

\bibliographystyle{JHEP}
\bibliography{references}

\providecommand{\href}[2]{#2}\begingroup\raggedright\begin{thebibliography}{10}

\bibitem{Eichmann:2016yit}
G.~Eichmann, H.~Sanchis-Alepuz, R.~Williams, R.~Alkofer and C.~S. Fischer,
  \emph{{Baryons as relativistic three-quark bound states}},
  \href{https://doi.org/10.1016/j.ppnp.2016.07.001}{\emph{Prog. Part. Nucl.
  Phys.} {\bfseries 91} (2016) 1}
  [\href{https://arxiv.org/abs/1606.09602}{{\ttfamily 1606.09602}}].

\bibitem{Ball:1980ay}
J.~S. Ball and T.-W. Chiu, \emph{{Analytic Properties of the Vertex Function in
  Gauge Theories. 1.}},
  \href{https://doi.org/10.1103/PhysRevD.22.2542}{\emph{Phys. Rev.} {\bfseries
  D22} (1980) 2542}.

\bibitem{Bali:2016lvx}
{\scshape RQCD} collaboration, G.~S. Bali et~al., \emph{{Direct determinations
  of the nucleon and pion $\sigma$ terms at nearly physical quark masses}},
  \href{https://doi.org/10.1103/PhysRevD.93.094504}{\emph{Phys. Rev.}
  {\bfseries D93} (2016) 094504}
  [\href{https://arxiv.org/abs/1603.00827}{{\ttfamily 1603.00827}}].

\bibitem{Gockeler:2010yr}
M.~G{\"o}ckeler et~al., \emph{{Perturbative and Nonperturbative Renormalization
  in Lattice QCD}}, \href{https://doi.org/10.1103/PhysRevD.82.114511,
  10.1103/PhysRevD.86.099903}{\emph{Phys. Rev.} {\bfseries D82} (2010) 114511}
  [\href{https://arxiv.org/abs/1003.5756}{{\ttfamily 1003.5756}}].

\bibitem{Arthur:2010ht}
{\scshape RBC, UKQCD} collaboration, R.~Arthur and P.~A. Boyle, \emph{{Step
  Scaling with off-shell renormalisation}},
  \href{https://doi.org/10.1103/PhysRevD.83.114511}{\emph{Phys. Rev.}
  {\bfseries D83} (2011) 114511}
  [\href{https://arxiv.org/abs/1006.0422}{{\ttfamily 1006.0422}}].

\bibitem{Leutnant:2018dry}
M.~Leutnant and A.~Sternbeck, \emph{{Quark-photon vertex from lattice QCD in
  Landau gauge}}, {\emph{PoS} {\bfseries Confinement2018} (2018) 095}
  [\href{https://arxiv.org/abs/1812.11131}{{\ttfamily 1812.11131}}].

\bibitem{Heatlie:1990kg}
G.~Heatlie, G.~Martinelli, C.~Pittori, G.~C. Rossi and C.~T. Sachrajda,
  \emph{{The improvement of hadronic matrix elements in lattice QCD}},
  \href{https://doi.org/10.1016/0550-3213(91)90137-M}{\emph{Nucl. Phys.}
  {\bfseries B352} (1991) 266}.

\bibitem{Skullerud:2000un}
J.~I. Skullerud and A.~G. Williams, \emph{{Quark propagator in Landau gauge}},
  \href{https://doi.org/10.1103/PhysRevD.63.054508}{\emph{Phys. Rev.}
  {\bfseries D63} (2001) 054508}
  [\href{https://arxiv.org/abs/hep-lat/0007028}{{\ttfamily hep-lat/0007028}}].

\bibitem{Oliveira:2018lln}
O.~Oliveira, P.~J. Silva, J.-I. Skullerud and A.~Sternbeck, \emph{{Quark
  propagator with two flavors of O(a)-improved Wilson fermions}},
  \href{https://arxiv.org/abs/1809.02541}{{\ttfamily 1809.02541}}.

\bibitem{Karsten:1980wd}
L.~H. Karsten and J.~Smit, \emph{{Lattice Fermions: Species Doubling, Chiral
  Invariance, and the Triangle Anomaly}},
  \href{https://doi.org/10.1016/0550-3213(81)90549-6}{\emph{Nucl. Phys.}
  {\bfseries B183} (1981) 103}.

\bibitem{Horsley:2015nae}
R.~Horsley, Y.~Nakamura, H.~Perlt, P.~E.~L. Rakow, G.~Schierholz, A.~Schiller
  et~al., \emph{{Improving the lattice axial vector current}},
  \href{https://doi.org/10.22323/1.251.0138}{\emph{PoS} {\bfseries LATTICE2015}
  (2016) 138} [\href{https://arxiv.org/abs/1511.05304}{{\ttfamily
  1511.05304}}].

\end{thebibliography}\endgroup

\end{document}